\documentclass[conference]{IEEEtran}
\IEEEoverridecommandlockouts
\usepackage[utf8]{inputenc}
\usepackage{cite}
\usepackage{blindtext}
\usepackage{multicol}
\usepackage{multirow}
\usepackage{graphicx}
\usepackage{indentfirst}
\usepackage{ragged2e}
\usepackage{float}
\usepackage{setspace}
\usepackage{notes2bib}

\usepackage{amsmath,amssymb,amsfonts}
\usepackage{algorithmic}
\usepackage{textcomp}
\usepackage{xcolor}
\usepackage{soul}
\usepackage{comment}
\usepackage{hyperref}

\usepackage{eurosym}
\usepackage{amstext} 
\DeclareRobustCommand{\officialeuro}{%
  \ifmmode\expandafter\text\fi
  {\fontencoding{U}\fontfamily{eurosym}\selectfont e}}

\def\BibTeX{{\rm B\kern-.05em{\sc i\kern-.025em b}\kern-.08em
    T\kern-.1667em\lower.7ex\hbox{E}\kern-.125emX}}

\begin{document}

\title{EVs and ERCOT: Future Adoption Scenarios and Grid Implications\\
\thanks{This research is partially supported by Carnegie Mellon University's Block Center for Technology and Society and by the Portuguese Foundation for Science and Technology through the research project ML@GridEdge (UTAP-EXPL/CA/0065/2021).}
}

\author{\IEEEauthorblockN{Kelsey Nelson}
\IEEEauthorblockA{\textit{Civil, Architectural and Env. Eng.} \\
\textit{The University of Texas at Austin}\\
Austin, USA \\
kelseynelson@utexas.edu}

\and
\IEEEauthorblockN{Pedro Moura}
\IEEEauthorblockA{\textit{ISR, Electrical and Computer Eng.} \\
\textit{University of Coimbra}\\
Coimbra, Portugal \\
pmoura@isr.uc.pt}

\and
\IEEEauthorblockN{Javad Mohammadi}
\IEEEauthorblockA{\textit{Civil, Architectural and Env. Eng.} \\
\textit{The University of Texas at Austin}\\
Austin, USA \\
javadm@utexas.edu }

}

\IEEEoverridecommandlockouts
\maketitle

\IEEEpubidadjcol

\begin{abstract} 

Electric vehicles (EVs) are becoming more commonplace in Texas, mainly due to their increasing attractiveness to consumers and pushes from the state's governing bodies to incentivize further adoption. Meanwhile, service from Texas's electric grid, ERCOT, has been seeing increases in power demand due to a growing population, increased air conditioning use, and pushes for electrification across other industries. The electrification of vehicles will only add to this demand increase. This paper focuses on evaluating different EV adoption, charging management, and policy scenarios, and how they will be expected to impact ERCOT, particularly with respect to peak demand increases. A strong increase in the peak demand can lead to challenges to keep the electrical grid's reliability, making it an important consideration for electrification in any sector. The anticipated impacts of EV adoption on peak demand are quantified using ERCOT's data on past generation and planned installations, the approximated effectiveness of EV incentives, EV charging profiles, and travel patterns. The results showcase the fact that the achievement of ambitious EV market share goals will be manageable on a statewide level regarding electricity supply into 2030, but will eventually necessitate ambitious charging management strategies in order to limit the EV fleet's potentially heavy impact on peak demand looking forward into 2050 and beyond.

\end{abstract}


\section{Introduction}

\subsection{Motivation} 

By the end of 2021, electric vehicles (EVs) contributed to less than $1\%$ of Texas’s light duty vehicle (LDV) fleet. While this only corresponded to about $120,000$ EVs on the state's roads, adoption rates have risen sharply within the last few years and this trend is expected to continue \cite{Registrations}. This trend has emerged in large part due to an increase in intrinsic demand from consumers as prices become more favorable, battery capacity expands, vehicle ranges increase, and the reduction of greenhouse gas (GHG) emissions and associated climate concerns become more important to consumers.

In addition to this intrinsic demand, United States federal and state policymakers have been looking to guide the country toward even higher EV adoption rates. This is in large part to help reach the Biden Administration's goal of $50\%$ GHG emissions reduction relative to 2005 levels by the year 2030 \cite{CarbonTarget}. Due to the much lower GHG emission rates of EVs when compared to traditional gasoline vehicles, the transportation industry has been targeted with a goal of achieving $50\%$ market share of EVs by 2030 \cite{EVTarget}. To reach this market share, there has been increased federal funding for EV tax credits and charging infrastructure installment, growing discussions and proposals of non-monetary incentives, and other incentivization occurring at the state level. 

While reaching this market share by the end of the decade would achieve the desired effect of decarbonizing a substantial portion of the transportation industry, there are concerns regarding its potentially negative implications on the electric grids servicing the EVs. As market share rises and more EVs are adopted in Texas, more and more miles traveled by its residents will be electrified, requiring supply from the Electric Reliability Council of Texas (ERCOT), the electric grid that services the vast majority of the state. ERCOT's peak demand has already been steadily increasing over the past decade, due to a variety of factors such as population growth, electrification of other appliances, and increasingly hot summers \cite{TexasWeather}. This means that any large scale electrification efforts should be carefully evaluated with their grid impacts in mind.

\subsection{Contribution}

This work evaluates how different scenarios for EV fleet growth could be expected to impact ERCOT. Scenarios were selected based on likelihood and relevance, taking into consideration factors such as currently implemented policy, market share goals, and practicality. The results were then used to create a policy recommendation framework in order to maximize the benefits of electrifying transportation in Texas without adding undue stress to the grid.

While similar analyses have been developed for large regions such as the United States and Europe as a whole, none have yet been conducted for Texas's regional grid \cite{NationalStudies}. Focusing on this region allows for the consideration of factors that can easily get lost on a national scale. Such factors include the state's unique travel patterns, electricity mix, weather conditions, shifting demographics, and existing policies and plans regarding EV adoption. Additionally, this study examines daily travel patterns in detail in order to account for variations in electricity use per mile for differing trip distances and daily miles traveled, which vary widely between different days of the week, seasons, and near holidays. These factors provide targeted insight and guidance for devising the most effective policy for ERCOT's unique challenges. 

\subsection{Organization}

The remainder of this paper is organized as follows. Section II presents the methods applied to the data used. Next, Section III presents the findings for the direct impact that electrification and charging patterns are expected to have on ERCOT. In Section IV, these findings are discussed in the context of relevant policy options for achieving different growth and market share scenarios. Finally, Section V presents the key takeaways from the achieved findings.  

\section{Datasets and Methods}

\subsection{Scope and Key Assumptions}

The methods, results, and discussion in this paper represent a broad analysis of all ERCOT serviced counties, making it suitable for informing decisions regarding statewide policy, operation, and infrastructure plans in Texas. It is assumed that Texas population growth will match the Office of The State Demographer's projections for 2030 and 2050, and that the LDV fleet will increase proportionally over time \cite{Population}. A steady vehicle turnover rate is also assumed.

At the time of this writing, ERCOT's own peak demand forecasts do not go beyond the year 2032. For peak demand estimates beyond this time frame, the effect of a fleet's EV percentage on peak demand in one year is assumed to be proportional to its effect in future years. This is because a predictable underlying driver for the change in the total size of a state's LDV fleet and electricity usage is its change in population. This relationship is given by (\ref{Find_PLI}), where $PLI$ is the peak load increase as a percentage of business as usual (BAU) demand, $\Delta EV_\%$ is the change in EV percentage, $EV_{load}$ is the charging demand from the EV fleet during the grid's time of peak load, $LDV_t$ is the number of light duty vehicles in Texas, and $PL$ is the ERCOT's BAU peak load.

\begin{equation}\label{Find_PLI}
PLI\% = \frac{\Delta{EV_{\%}}\times {EV_{load}}\times {LDV_{tl}}}{PL}
\end{equation}

If the LDV fleet and ERCOT load both scale according to population, population growth will not change the peak load increase percentage. This means that after projecting the impacts of an incentive set into a year of interest and finding that year's estimated EV percentage, that percentage can still be evaluated using older data to find the peak demand increase as a percentage. 

It is also important to note that ERCOT predicts that there will be $1$ million EVs on the road in Texas by 2030 \cite{ERCOTProjections}. This estimate is quite conservative relative to other projections and targets, equating to about one third of what a $50\%$ market share would achieve and a similar fraction of predictions from the International Energy Agency \cite{IEA}. Such a small fleet's influence on peak demand within this time horizon would be under $2\%$, meaning that accounting for ERCOT's assumed EV fleet size in their own peak demand projections provided an insignificant change for the results. Due to this negligibility, ERCOT's projections were used as is (without subtracting the amount they attributed to EVs) in order to find peak demand increases from EV percentage increases.

Due to uncertainties regarding the advancement of EV technology and potential changes in travel patterns, daily miles traveled and mileage (miles/kWh) are assumed to remain consistent on a per-person basis. The mileage used takes into account how much of each trip was likely spent driving on highways vs residential roads. This is because longer trips are typically taken via highway at a faster speed, which decreases mileage for EVs, while shorter trips are typically taken via residential roads at slower speeds with higher mileage \cite{Mileage3}. Lastly, due to uncertainties with the advancement of electricity generation technology and changes in weather patterns, is assumed that wind and solar farms on ERCOT will have similar hourly and seasonal output profiles.

\subsection{Data Sources and Methods}

\begin{figure}[H]
\centering
\includegraphics[scale=.33]{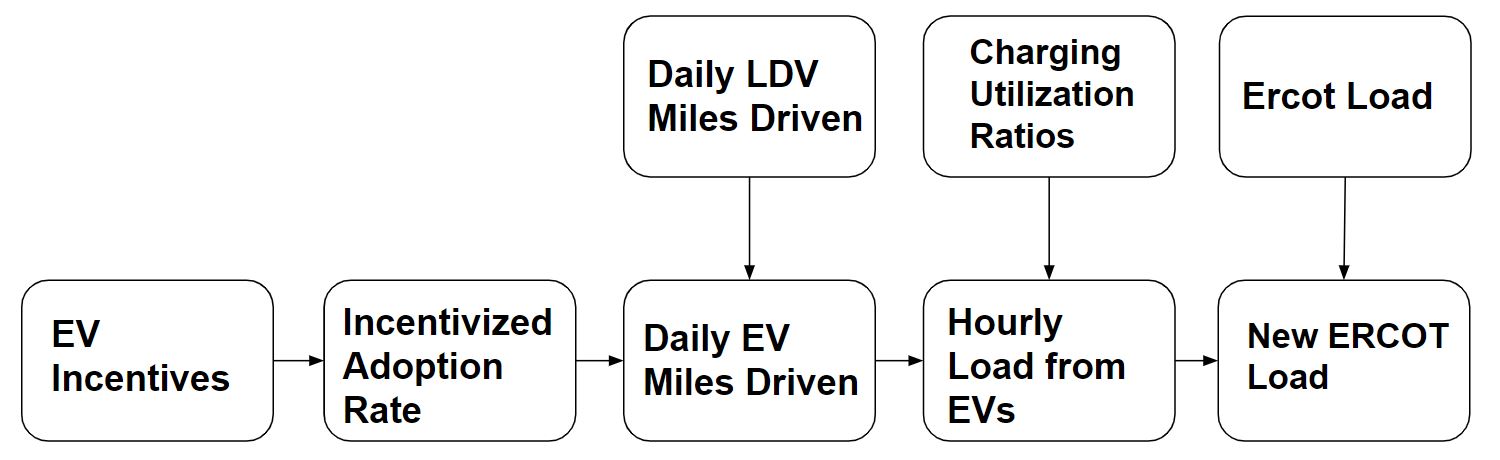}
\caption{Overview for Finding incentives' resultant impacts on ERCOT}
\label{Flowchart}
\end{figure}

Fig.~\ref{Flowchart} provides a visualization of the broad steps taken to find how a specific set of incentives may affect EV adoption, and by extension, ERCOT.

First, the incentives in question are quantified in units that are compatible with the available data from predictive models regarding incentive effectiveness. They are then used to estimate how they would result in a new, incentivized EV percentage. This percentage is then applied to the daily distance traveled within ERCOT serviced counties to give the total daily distance traveled by \cite{BTS2022}. These distances are then converted into a daily electricity demand (in $kWh$) from the average specific energy demand for EVs (in $kWh/km$) \cite{Mileage2}. In order to convert the load from these daily demands to load from hourly demands, data from \cite{GridPIQ} was used, which provides average distributions for EV charging profiles by hour. These distributions were measured from several charging stations and households across Texas and are presented on a per-vehicle basis, meaning that they are scaleable to EV fleet size. Once the daily load was distributed by hour, this data was paired with ERCOT hourly load data in order to find how the addition of these EVs would increase peak load. 

The evaluations and findings presented in this paper rely in part on differences between BAU scenarios (where EV growth is only driven by intrinsic demand) and their counterparts (where EV adoption is manipulated through incentives). Trends for EV adoption from the past 10 years in the United States were analyzed in \cite{NBERw28933}, which outlines three BAU scenarios dependent on the level of intrinsic demand from the US consumer base as a whole for EVs. These scenarios were used in this paper to forecast adoption rates for EVs through 2035, which each have their own year on year (YoY) adoption rate decline: low growth corresponds to a $15\%$ YoY growth rate decline, medium growth corresponds to a $10\%$ YoY growth rate decline, and high growth corresponds to a $5\%$ YoY growth rate decline. This provides useful base cases to then compare incentivized adoption rates and market share achievement scenarios.

The effect of different incentives on adoption rates is discussed in \cite{Narassimhan_2018}, which sought to quantify the average effect of commonly proposed or used incentives on new EV registrations by developing coefficients that correspond to the percent change in registered vehicles. Of particular interest for this research was the effect of tax credits, rebates, and high occupancy vehicle (HOV) lane access, all of which have been implemented at some level within the United States. With these coefficients, new incentivized adoption rates were developed for different scenarios to examine the EV percentage increase for a year of interest.

Once the incentives' impacts on EV percentage were evaluated, their resultant changes in electricity demand from their use could be determined. This first required the examination of vehicular travel patterns. In \cite{BTS2022}, datasets for daily travel over all of Texas broken down to a county level are provided. Because this dataset was created from GPS tracking, it only logs each person trip, and not each vehicle trip, which is of more immediate concern because they can directly show the percentage of the distance traveled that could potentially be electrified within the LDV fleet. 

Hence, this was supplemented with data from \cite{FHWA} and \cite{BTS2011}, which provide information on how many Americans choose to fly versus travel by car for given trip distances, how many take public transportation, and the average party size for trips by distance and purpose. Data prior to 2020 was used to avoid the impact of COVID-19 lockdown measures on daily commutes and vacations. This allowed for the creation of multipliers for each trip distance category to convert its total person trips to vehicle trips. This conversion eliminates the person trips not taken by vehicle and accounts for carpooling. These multipliers are presented in Tab.~\ref{VT_Multipliers}.

\begin{table}[H]
\caption{Vehicle trips per person trips multipliers by distance}
\label{VT_Multipliers}
\centering
\begin{tabular}{|c|c|}
\hline
Trip Distance & Multiplier [VT/PT] \\
\hline

$<$5 km ($<$ 3 miles) & 0.684 \\
\hline
5-160 km (3-100 miles) & 0.922\\
\hline
160 - 400 km (100-250 miles) & 0.515\\
\hline
400 - 800 km (250-500 miles) & 0.513\\
\hline
800+ km (500 + miles) & 0.508\\
\hline
\end{tabular}
\end{table}

Once each day's miles traveled were converted to LDV miles traveled, \eqref{DailykWh} was used to find how this would translate to a daily electricity usage (in $kWh$). Where ${EV_{\%}}$ is the EV percentage, ${Miles_{LDV}}$ is the daily miles traveled by the LDV fleet each day, and ${EV_{mileage}}$ is the average mileage of an EV (in $kWh/mi.$), which are used to find $Euse_{EV}$, which is the daily EV electricity usage (in $kWh$) attributable to an LDV fleet size with a given EV percentage, \cite{Mileage}. 

\begin{equation}\label{DailykWh}
Euse_{EV} = {Miles_{LDV}}\times {EV_{\%}}\times {EV_{mileage}} 
\end{equation}

The travel data was evaluated on a day by day basis in order to account for variations in travel patterns due to different weekdays, seasons, and holidays. This electricity consumption (in kWh per day) was then distributed by the hour according to Grid PIQ’s recorded charging profiles for EVs in Texas \cite{GridPIQ}. After finding the new hourly loads for the entirety of a sample year, the estimated change in peak load could be found for a given EV percentage.

The relationship between changes in the fleet's EV percentage and peak load increase was evaluated to find the average peak load increase factor (PLIF) determined by \eqref{PLIF}. This factor describes the average ratio between a one point increase in EV percentage and a one percentage point increase in peak load. 

\begin{equation}
PLIF_\% =\frac{PeakLoadIncrease_\%}{EV_\%}
\label{PLIF}
\end{equation}

In order to evaluate the ability of charging management to mitigate peak load increase attributable to EV adoption, charging profiles were used with estimates from The Alternative Fuels Data Center (AFDC), which provides a tool that simulates how changes in charging infrastructure, technology, and incentives can impact the overall charging profile of an EV fleet \cite{AFDC}. The charging management strategies that these profiles are able to account for are the implementation of delayed charging, increased workplace charging availability, and even-spread charging.

For projections beyond 2030 and into 2050, an altered approach was used. Because the most notable piece of legislation regarding EV adoption, the Inflation Reduction Act of 2022, will only be in place through 2031, further projections did not rely on the estimated impact of specific incentive sets as it is too difficult to know what public attitude, policy, and fleet makeup will look like that far into the future \cite{IRA}. Instead, market share achievements were evaluated on their own; For a 2030 market share benchmark, a corresponding 2050 market share benchmark was estimated using projections from \cite{USDrive}, which outlines projections from 2030 based on different EV fleet growth scenarios into 2050. Market share benchmarks are modeled as being approached linearly over time, and the total number of registered EVs at the end of year $n$ is determined using \eqref{EV_total_N}. Where $EVt$ is the total number of EVs on the road, $MS$ is EV the market share, and ${LDVt}$ is the total number of registered light duty vehicles.

\begin{equation}\label{EV_total_N}
EVt_{n}=\frac{15}{16}{EVt_{n-1}}+MS_{n}*\frac{LDVt_{n}}{16}
\end{equation}

Only 15/16ths of the previous year’s EV fleet is carried onto the next year because 16 years is the average length of time that LDVs are used before being discarded or scrapped in the United States, so this accounts for EVs going out of commission over time (Keith et al, 2019). This is also why the LDV fleet is divided by 16, to reflect the approximate number of new vehicles being purchased as their replacement each year. Though they are more uncertain, these distant market share scenarios are still important to consider because as EV adoption continues over time, their impact on ERCOT will become increasingly significant. 

\section{Results}
The ratio between a one point increase in the fleet’s EV percentage and the percentage point increase in ERCOT peak demand was found to be $0.41$ \eqref{PLIF_percent}.

\begin{equation} \label{PLIF_percent}
PLIF(\%) =0.41
\end{equation}

This value for PLIF was found with BAU charging patterns. In Texas, peak demand occurs on late summer afternoons around $6 pm$. As it stands, the vast majority of EV owners are middle to upper-class people, who are likely to hold day jobs that also end around this time. Because of this, most EV owners tend to begin charging their vehicles when they return home from work while demand is already peaking, which further exacerbates the problem of load concentration in the late afternoon.

\begin{figure}[htbp]
\centering
\includegraphics[trim=10 20 20 70,clip, width=0.48\textwidth] {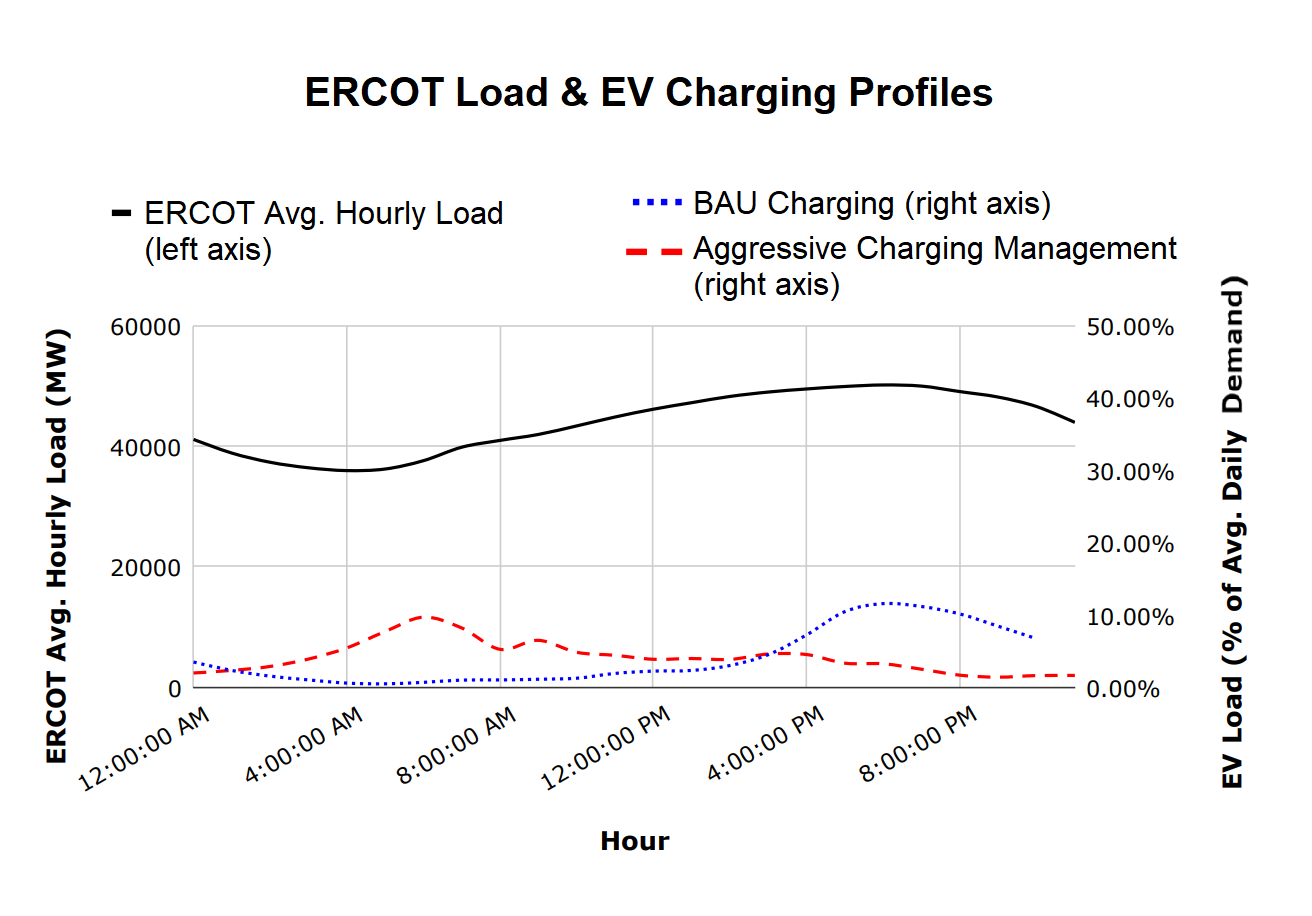}
\caption{ERCOT average hourly demand (left axis) and the EV Load resulting from differing charging profiles (right axis)}
\label{ChargingOptions}
\end{figure}

The use of charging management strategies will be an important tool for mitigating the peak demand increasing effect of EV charging. By utilizing delayed charging, slower, more distributed charging, and providing more chargers and public and workplace buildings, EV owners would be more likely to charge and store this electricity in their vehicles during other hours of the day. The result of how this could cause the EV charging load to shift away from ERCOT's peak load is shown in Fig.~\ref{ChargingOptions}. Shifting to this charging profile would lower the peak demand increase from EVs to as low as $35\%$ of what it would have been without charging management.

\section{Discussion}

In August 2022, the congressional bill H.R.5376, titled Inflation Reduction Act (IRA), was passed with measures for the federal government to continue incentivizing the adoption of EVs. Notably, it introduced a tax credit of up to $\$4,000$ for the purchase of a used EV and continued an already in place $\$7,500$ tax credit, first implemented in 2008, for new EVs with some alterations \cite{2008}. This credit is now in place as the "Clean Vehicle Credit" \cite{IRA}. The Clean Vehicle Credit differs from the 2008 EV tax credit in that it implements a requirement for the vehicles to be manufactured in the US and one that will phase in through 2028 requiring the battery and mineral components to be sourced from the United States. 

Though these sourcing and manufacturing requirements have been introduced, it nixes the $200,000$ vehicle per manufacturer limit that was present in the 2008 credit. This means that manufacturers that have already sold $200,000$ EVs, such as Tesla and General Motors, can now qualify again. In Texas, Tesla alone accounts for over half of the EVs on the road, meaning the Clean Vehicle Credit will likely be more effective than its 2008 counterpart when it comes to getting more EVs on the road in  \cite{Registrations}.

Current policies available to Texans include this federal tax credit of $\$7,500$ in the Inflation Reduction Act and a $\$2,500$ rebate available through the state. There is also a plan to implement charging stations along major highways placed no more than $50$ miles apart. Tab.~\ref{Incentives_Results} presents the resulting impact on ERCOT in 2030 for low, medium, and high growth scenarios with these incentives, both with and without charging coordination. 

\begin{table}[htbp]
\caption{Peak load increase in 2030 due to current policy on ERCOT under different growth and charging scenarios} 
\label{Incentives_Results}
\begin{center}
\begin{tabular}{|c|c|c|}
\hline
Growth & Unmanaged & Managed\\
Scenario & Charging & Charging\\
 \hline
 Low  & 1.74\% & 0.609\% \\
 \hline
 Medium & 2.04\% & 0.714\% \\
 \hline
 High & 2.47\% & 0.87\% \\
 \hline
 \end{tabular}
 \end{center}
\end{table}

With effective demand response controls, the peak demand impact on ERCOT could be limited to less than $1\%$ by 2030. ERCOT projects a maintaining reserve margin for over $30\%$ of peak demand through this time frame, meaning that even a high growth, unmanaged scenario for current policy options would not threaten grid conditions \cite{ERCOTPeak}.

The scenario examined above results in a $30\%$ market share for high growth, which still falls short of the $50\%$ market share goal despite assuming the most optimistic levels of intrinsic demand from the consumer.
If incentives become great enough to result in a growth scenario where the nation reaches $50\%$ market share by 2030, the grid impacts become more evident, especially as this growth is extended to the year 2050. These results, which are obtained through \eqref{EV_total_N} and \eqref{Find_PLI} are summarized in Tab.~\ref{Market_Share}.

\begin{table} [htbp]
\caption{Peak load increase associated with Biden administration market share goals and growth}
\label{Market_Share}
\begin{center}
\begin{tabular}{|c|c|c|c|}
\hline
Year & Market & BAU & Managed\\
  & Share  & Charging & Charging \\
 \hline
2030 & 50\% & 4.6\% & 1.61\% \\
 \hline
2050 & 90\% & 32.8\% & 11.5\% \\
 \hline
 \end{tabular}
  
 \end{center}
\end{table}

Though ERCOT does not have grid projections into the year 2050, sustained growth from a 2030 market share achievement would result in a $32.8\%$ increase in peak demand, which given historical ERCOT reserve margins would result in blackouts and brownouts \cite{ERCOTPeak}.

\section{Conclusions}

Regarding policy, this paper shows that achieving a $50\%$ market share goal will necessitate further incentivization, as current incentives only show the potential to drive market share to about $30\%$. Though there are strong incentives in place, they will only be at their most effective if intrinsic demand rises among the general population, battery component sourcing is able to shift quickly enough to the domestic market for tax credits to be claimable, and if charging infrastructure is made widely available. 

This paper does not provide analyses at the local level, which could be of interest to city planners and funding allocation for charging infrastructure. It also does not tackle factors that are surrounded by inherent uncertainty, such as future improvements in EV mileage, changes to ERCOT's planned installation, or changes in weather patterns that may impact peak demand. These factors would certainly be of interest for future or continued research. Despite these uncertainties, however, a broad state level overview is provided regarding what changes can be expected for the achievement of ambitious EV market share goals and what strategies may become increasingly important to consider in order to deal with the negative aspects of these changes.

Without charging coordination, a $50\%$ market share achievement by 2030 could increase peak load by about $4.6\%$. This peak load increase could grow to over $30\%$ by 2050. Without additional peak load capacity on standby, this is high enough to threaten brownouts or blackouts. Because of this, the continued implementation of public charging, workplace charging, and delayed home charging is critical to avoid the overlap of charging peak demand and the rest of ERCOT peak demand. This could result in a reduction of peak demand to as low as one third of what it would be with BAU charging patterns. This paper demonstrates that with charging management, supplying sufficient electricity to a rapidly growing EV fleet would be significantly less disruptive to ERCOT, making LDV electrification in Texas a viable tool for reducing GHG emissions with proper long term planning and resource allocation. 

\vspace{.2cm}
\bibliographystyle{IEEEtran}
\bibliography{main}

\end{document}